\documentclass[fullpage,10pt,doublecolumn]{article}
\usepackage{array}
\usepackage{amsmath, amsthm, amssymb} 
\usepackage{algpseudocode}
\usepackage{algorithm}
\usepackage{multirow}
\usepackage{bigstrut}
\usepackage{lastpage}

  \usepackage[dvips]{graphicx}
  \graphicspath{{./figures/}}
  \DeclareGraphicsExtensions{.eps}

\def\x{{\mathbf x}}

\begin{document}

\title{Data-driven Approaches for Social Video Distribution}

\author{
Zhi Wang\\
       {Tsinghua University}\\
}

\maketitle

\begin{abstract}
	
	The Internet has recently witnessed the convergence of online social network services and online video services: users import videos from content sharing sites, and propagate them along the social connections by re-sharing them. Such social behaviors have dramatically reshaped how videos are disseminated, and the users are now actively engaged to be part of the social ecosystem, rather than being passively consumers. Despite the increasingly abundant bandwidth and computation resources, the ever increasing data volume of user generated video content and the boundless coverage of socialized sharing have presented unprecedented challenges. In this paper, we first presents the challenges in social-aware video delivery. Then, we present a principal framework for data-driven social video delivery approaches. Moreover, we identify the unique characteristics of social-aware video access and the social content propagation, and closely reveal the design of individual modules and their integration towards enhancing users' experience in the social network context.

\end{abstract}

\section{Introduction} \label{sec:intro}

In the past decade, we have witnessed the fast evolution toward new generation networked multimedia processing and sharing in the Web 2.0 era. The users are now actively engaged to be part of a social ecosystem, rather than passively receiving media content. The revolution is being driven further by the deep penetration of 3G/4G wireless networks and smart mobile devices that are seamlessly integrated with online social networking and media sharing services. 

Despite the increasingly abundant bandwidth and computation resources, the ever increasing data volume of user generated video content and the boundless coverage of socialized sharing have presented unprecedented challenges to both content and network service providers. The highly diversified content origins and distribution channels further complicate the design and management of online video sharing systems. This paper identifies the key issues and presenting our data-driven solution toward this promising research direction.

\subsection{Background and Challenges in Social Video Delivery}

Online social network services connect users through ``friendship'' (\emph{e.g.}, Facebook), ``following'' relationship (\emph{e.g.}, Twitter), or professional connections (\emph{e.g.}, LinkedIn). Such applications have successfully changed how people are connected to each other, and how they share information, including video. We have recently seen the convergence of online social network services and online video services: users can ``import'' videos from video sharing sites to online social networks, and make the videos propagate along the social connections by \emph{re-sharing} them. The social behaviors have dramatically reshaped how videos are disseminated to users: \emph{people are now receiving videos from friends directly}. For example, the online video clip ``Gangnam Style'', attracted over $1$ billion views in $6$ months after it is uploaded, due to its propagation over popular online social networks including Twitter and Facebook. Today, $500$ years of video are watched and shared every day by Facebook users, and over $700$ videos are shared on Twitter each minute~\cite{youtube-stat}.

Conventional video delivery strategies, \emph{e.g.}, the original Client/Server streaming, IP and Application-Layer multicast / peer-to-peer, and Content Delivery Networks (CDNs), mainly focus on improving the network delivery performance, so as to meet the increasing scale of video requests. They have generally assumed that the content comes from centralized providers and the users only \emph{passively} receive the contents~\cite{zhi-tmm2013}. For videos shared over social networks, however, the access patterns are much more dynamic, being affected by individuals and their activities during propagation. With information propagation, the scale/coverage in the social network context can be much larger and broader, making the system design much more involving.

\begin{figure*}[!t]
	\centering
		\includegraphics[width=0.85\linewidth]{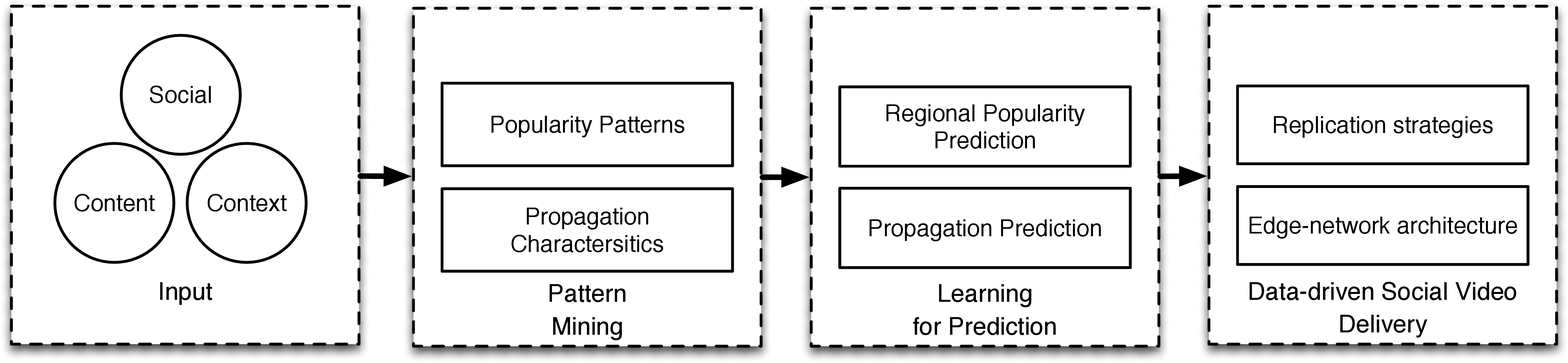}
	\caption{Framework of social-aware video content delivery.}
	\label{fig:framework}
\end{figure*}

More specifically, we are facing the following challenges in social video delivery.

First, users instead of content providers determine how videos reach each other. In an online social network, contents are generated, propagated, and disseminated by users. In 2014, YouTube reported that over $100$ hours of video clips were uploaded by users in every minute. The content delivery systems thus have to distribute a much larger volume of user-generated videos than what has been handled by conventional content providers ever \cite{cha2007tube}. Second, the users share videos through social connections, and they tend to receive videos from their friends. As such, a provider no longer has tight and centralized control over the dissemination of contents.

Second, dynamical content propagation. Social propagation is affected by a combination of factors, including the social topology, user behaviors, and the inherent content characteristics, to name but a few~\cite{zhi-tpds-enhancing}. Given the many influential factors, propagation is highly dynamical, and traditional content delivery strategies generally lack prediction tools to infer such inherent dynamics.

Third, change of social content popularity. The change of the content origin and the social propagation also change the popularity distribution of videos. On one hand, the overall skewness of social video popularity distribution has been amplified; on the other hand, certain unpopular videos can be re-identified and then shared among users with close social relations. Video popularity is a key factor in designing and optimizing video delivery systems, and the change thus has strong implications. For example, it has been observed that the request hit ratio can be degraded by over $70\%$ when traditional cache strategies were used to handle online social contents~\cite{misloverethinking}.

\subsection{Data-driven Social Video Delivery Framework}

These challenges demand a joint study on user behaviors, social video popularity, and social propagation to enhance the video content delivery in the social network context. To this end, Fig.~\ref{fig:framework} illustrates a general framework of social-aware video content delivery. This integrated framework takes the social, content and context information as the input to mine the popularity patterns and propagation characteristics. A machine learning model is designed to predict the evolution of social video popularity and the propagation patterns (\emph{e.g.}, the size of a social cascade). The prediction reveals how videos would be shared at different regions, and the network resources can then be adaptively allocated among these regions. Content replication and caching strategies will be incorporated as well to further improve the sharing efficiency~\cite{torres2011dissecting}.

We next check the unique characteristics of social-aware video access and the social content propagation, and closely examine the design of individual modules and their integration in the framework.

\section{Social Video Popularity}

We start from the popularity of videos propagated through online social networks, including the distribution, evolution, and effective prediction.

\subsection{Social-Aware Popularity Distribution}

\begin{figure}[t]
	\centering
		\includegraphics[width=0.8\linewidth]{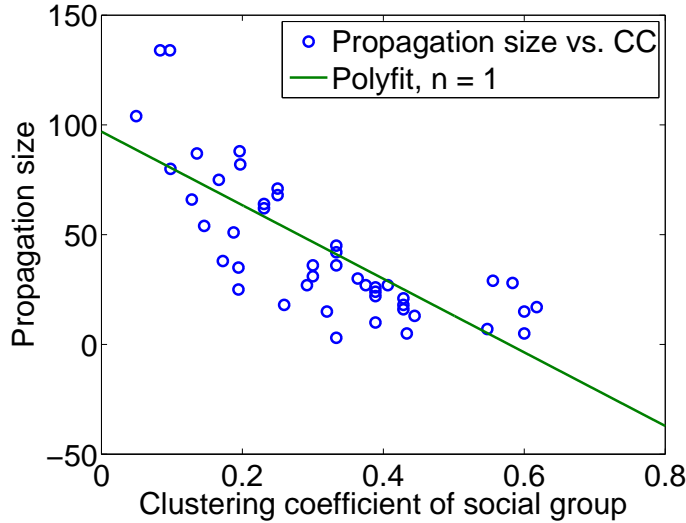}
	\caption{Unpopular videos tend to be shared in small social groups with membered socially connected.}
	\label{fig:view-vs-cc}
\end{figure}

Since there is a large fraction of unpopular social videos, we further investigate in which types of social groups these unpopular videos propagate, using a Weibo (a Twitter-like social network) dataset that provides traces recording $400,000$ videos shared by users in $1$ month. By randomly sampling $50$ videos with different propagation size (the number of users involved in a video's propagation), we explore the correlation between the propagation size and the clustering coefficient (larger clustering coefficient indicates close connection between members) of the social group formed by the users involved in the propagation. In Fig.~\ref{fig:view-vs-cc}, each sample illustrates the video propagation size versus the clustering coefficient of the corresponding social group. We observe a relatively strong correlation between the propagation size and the clustering coefficient; that is, unpopular videos tend to be shared among small social groups that are closely-connected (socially).

\subsection{Social Popularity Prediction}

The above observations suggest that content delivery mechanisms need to be substantially revised,  and the social-related factors (\emph{e.g.}, the total number of previous views and the video age) will play important roles~\cite{borghol2012untold}. In particular, the popularity prediction has to jointly consider both \emph{accuracy} and \emph{timeliness} of the prediction results. 

As shown in~\cite{xu2014forecasting}, each video can be assigned with an index $k\in\{1,2,...,K\}$ according to the absolute time $t_{init}^{k}$ when the video is initiated. Once a video is shared, it will be propagated through the social network for some time duration. A video has an age of $n\in\{1,2,...\}$ periods if it has been propagated through the social media for $n$ periods. In each period, the video is further shared and viewed by users depending on the sharing and viewing status of the previous period. The propagation characteristics of video $k$ up to age $n$ are captured by a $d_{n}$-dimensional vector $\x_{n}^{k}\in\mathcal{X}_{n}$ which includes information such as the total number of views and other contextual information, such as the characteristics of the social network over which the video propagates.

The popularity prediction can then be formulated as a multi-stage sequential decision and online learning problem \cite{xu2014forecasting}. Using multi-level popularity prediction in an online fashion, such a social-aware prediction outperforms existing view-based approaches by more than $30\%$ in terms of prediction reward (a tradeoff between the popularity prediction accuracy and the timeliness).

\section{Dynamic Social Video Propagation}
\label{sec:propagation}

The popularity reflects the macroscopic aggregated views of the videos shared. We next have a closer look at the social propagation process for videos, which determines how individual videos reach different users in online social networks.

\subsection{Propagation Patterns}

The generation and re-share of a social video typically form a propagation tree, rooted at the user who generates the video or initiates the sharing (referred to as the \emph{initiator} or \emph{root}). We refer to the users who re-share the video as \emph{spreaders}, and the users who receive the shared video as \emph{viewers} (or \emph{receivers}). A video's popularity can then be calculated as the sum of its spreaders and receivers.

Large-scale measurement studies have also discovered interesting locality patterns in the propagation structures~\cite{zhi-acmmm2012}.

\begin{figure}[t]
	\centering
		\includegraphics[width=0.81\linewidth]{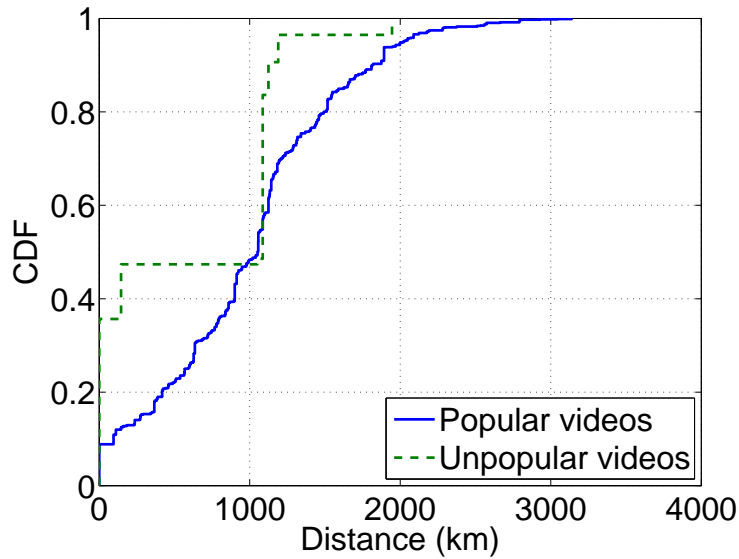}
	\caption{CDF of distance between users that are sharing in the same propagation of social videos.}
	\label{fig:distance-cdf}
\end{figure}

$\rhd$ \emph{Geographical locality}: A large fraction of the videos are shared between users who are geographically close to each other. In Fig.~\ref{fig:distance-cdf}, we plot the CDF of distances between users who join the social propagation of the same video in the online social network. Different curves are for videos with different popularities: (1) Popular videos: videos whose popularity is in the top $2\%$; (2) Unpopular videos: videos whose popularity is in the bottom $30\%$. We observe that different from traditional video consumption, unpopular social videos tend to be shared in local regions, where users are nearby to each other. For example around $40\%$ of the distances between users sharing the same unpopular videos are close to $0$km.

\begin{figure}[t]
	\centering
		\includegraphics[width=0.8\linewidth]{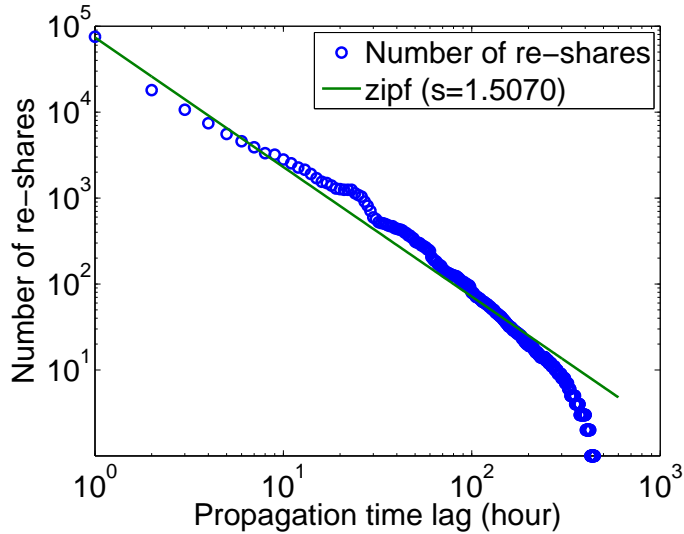}
	\caption{Number of re-shares versus the time lag.}
	\label{fig:propagation-interval}
\end{figure}

$\rhd$ \emph{Temporal locality}: In the online social network, the users are more likely to re-share new video contents, \emph{i.e.}, videos that are recently imported or re-shared. Fig.~\ref{fig:propagation-interval} illustrates the number of re-shares of a video in a timeslot ($1$ hour) versus the time lag since the propagation. We observe that most of the re-shares happen in the recent hours, and the re-share number against the time lag follows a zipf-like distribution with a shape parameter $s=1.5070$. More than $95\%$ of the re-shares happen within the first 24 hours. This indicates that in social video sharing, users' behaviors are highly crowded around the time point when it is imported.

\subsection{Propagation Model}

Given the propagation patterns and the different types of nodes, an \emph{SIR (Susceptible-Infectious-Recovered) model} can be used to model social video propagation. It considers a fixed population with three compartments: \emph{susceptible}, \emph{infectious}, and \emph{recovered}. Li \emph{et al.}~\cite{li2014understanding} particularly investigate an extended epidemic model to capture the video propagation in online social networks:

Initially, a user shares this video from an external video sharing site and this initiator becomes \emph{infectious}. All other users in the social network are \emph{safe} except the friends of the initiator. The shared video appears in the news feed of the initiator's friends and thus they become \emph{susceptible}. After a while, these friends log into the social network gradually and decide whether to watch the video (\emph{infected}) or not (\emph{immune}). For those \emph{infected} users, they will decide whether to share after watching the video. They become \emph{recovered} if they choose not to share. They will become \textsf{Infectious} if they choose to share. Again, these \emph{infectious} users will make their friends who are on \emph{safe} stage become \emph{susceptible}.

Based on the propagation model, the parameters can be trained using real propagation traces, and the connection between video access and social activities is created for content delivery strategies.

\section{Data-driven Social Video Delivery}

Given the unique characteristics of the social videos, a series of solutions have been proposed in the literature, addressing the challenges in different aspects.

\subsection{Propagation Prediction based Replication}

A significant amount of efforts have been devoted to utilize social information to enhance the strategies in the traditional content delivery (\emph{e.g.}, CDN, Peer-to-Peer Network). Wang \emph{et al.}~\cite{zhi-acmmm2012} proposed a hybrid CDN and P2P architecture for social video distribution, where the CDN servers can support the time-varying bandwidth and storage allocations requested by different regions, while peers are able to help contribute to each other in similar social groups.

\begin{figure}[t]
	\centering
		\includegraphics[width=0.7\linewidth]{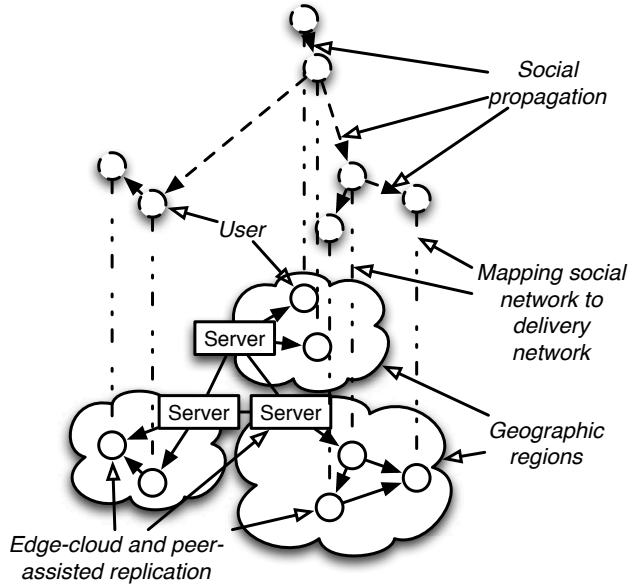}
	\caption{Edge-cloud and peer-assisted social replication.}
	\label{fig:architecture}
\end{figure}

In this replication architecture, two overlays exist in the social-aware replication design (as illustrated in Fig.~\ref{fig:architecture}): (1) \emph{social propagation overlay} based on the social graph, which determines the video propagation among friends, \emph{i.e.}, users after generating a video, can share the video with their direct friends, who will further re-share the video to more people, and (2) \emph{delivery overlay} which determines how video contents are delivered from edge-cloud servers to users or among themselves in a peer-assisted paradigm.

\emph{Geographic Influence Index}: The replication is designed to take social propagation into account.  When performing video replication, we need to find out the videos that may propagate to more regions in the future. We design a geographic influence index in the region prediction for that, as below:
$$
	g_v^{(T)} = c_1  \log(c_2 s_v^{(T-1)}),
$$
where $s_v^{(T-1)}$ is the propagation size of the propagation tree of video $v$ in timeslot $T-1$. $g_v^{(T)}$ approximates the number of regions where the video will propagate to in the future time window, according to our measurement studies. Intuitively, a video should be replicated to more regions when the predicted number of regions involved in the propagation is larger than the number of regions it has already been replicated to ($\Theta$).
To proactively replicate social videos, a video with a geographic influence index $g_v^{(T)} > \Theta$ should be replicated to more regions to serve users locally. Parameters $c_1$ and $c_2$ are selected to fit the measurements. Based on the geographic influence index, we can predict whether the regions where the video has been replicated are enough.

\subsection{Mobility and Propagation Prediction based Caching}

Crowdsourced caching by replicating bandwidth-intensive videos on the \emph{edge-network} devices (\emph{e.g.}, users' smartphones) to let users serve each other is promising. Previous studies have demonstrated that such device-to-device (D2D) content sharing is possible when users are close to each other, and the contents to be delivered by users are delay tolerant \cite{wang2014toss}. However, in traditional D2D content sharing, a user broadcasts generated contents or reshared contents to a set of random users that are close to it. As a result, all contents are disseminated to users in the same way (\emph{e.g.}, random flooding), causing the following problems: (1) In greedy flooding, smart devices in edge networks have to spend expansive power to cache and relay excessive contents. As the number of user-generated social contents is increasing, such mechanism is inherently non-scalable; (2) Social videos have heterogeneous popularity, while the conventional approaches treat them all the same, resulting in wasted resource to carry unpopular contents. (3) Due to the dynamic mobility patterns, users may not be able to fetch contents timely, resulting in poor quality of experience.

\begin{figure}[t]
	\centering
		\includegraphics[width=\linewidth]{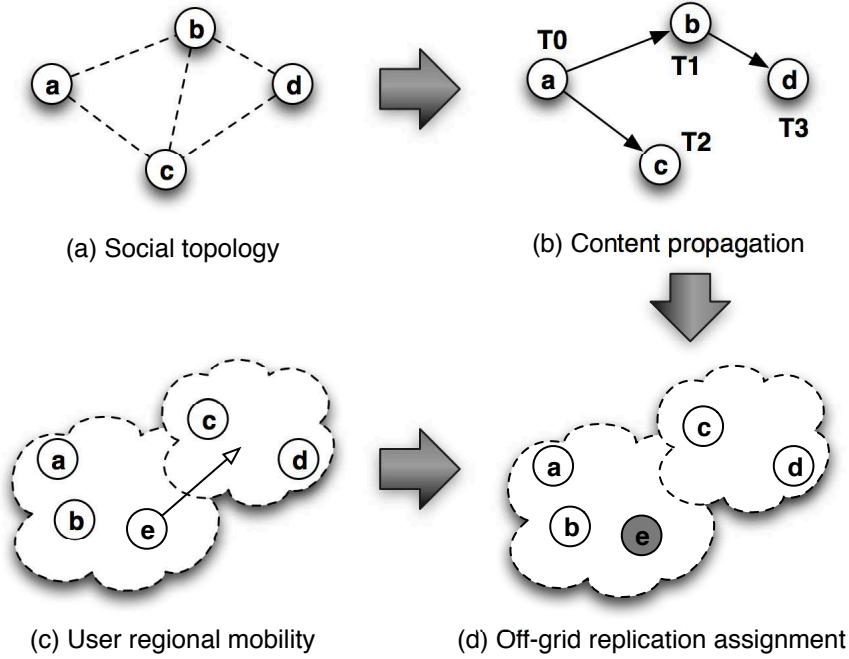}
		\caption{Crowdsourced replication affected by social topology, content propagation and user regional mobility.}
	\label{fig:user-topologies}
\end{figure}

To address these issues, a joint propagation- and mobility-aware crowdsourced replication strategy can be developed based on social propagation characteristics and crowd mobility patterns in edge-network \emph{regions}, \emph{e.g.}, an area of hundred-meter range where users can move across. As illustrated in Fig.~\ref{fig:user-topologies}, using the social graph and propagation patterns, we first estimate how contents will be received by users, and we then predict to which regions the users will be moving and how long they will stay. Instead of letting contents flood between users that are merely close to each other, we disseminate social videos according to the influence of users and the propagation of videos. In this example, as user $e$ -- while not a friend of any other user -- is moving to the region where user $c$ and $d$ are. Thus, $e$ will be selected to replicate the content generated by user $a$, and both user $c$ and $d$ will receive the content shared by user $a$ in the social propagation at times $T2$ and $T3$, respectively.

\section{Summary and Future Direction} \label{sec:conclusion}

With the advances of online social networking, users, instead of content providers, determine how videos reach each other. Such key characteristics of networked videos as the popularity distribution and its evolution have been strongly affected by the social behaviors of the users, challenging the traditional content delivery that consider users as passive consumers only. In this paper, we identified the unique patterns and characteristics of social video propagation and content popularity through large-scale trace data analysis. We demonstrated a series of strategies of great potentials, including content replication, crowdsourced content caching, and network resource allocation. This new and promising research area opens many challenging issues to be addressed in the near future, and we believe that a data-driven framework design will be the key. Within this framework, deeply understanding on the user behaviors, the social propagation structures, the knowledge of content characteristics and context information, as well as social-relationship-based collaborative content sharing mechanisms will all play important roles.

\bibliographystyle{plain}
\bibliography{mylib}

\end{document}